\documentclass[review]{elsarticle}

\journal{Journal of \LaTeX\ Templates}

\begin{document}

\begin{frontmatter}

\title{Bootstrap validation of links of a minimum spanning tree}


\author[address1]{F. Musciotto}

\author[address1]{L. Marotta}

\author[address1]{S. Miccich\`e}

\author[address1,address2,address3]{R.N. Mantegna\corref{mycorrespondingauthor}}
\cortext[mycorrespondingauthor]{Corresponding author}
\ead{rosario.mantegna@unipa.it}



\address[address1]{Dipartimento di Fisica e Chimica, Universit\`a degli Studi di Palermo, Viale delle Scienze, Ed. 18, I-90128 Palermo, Italy}

\address[address2]{Department of Computer Science, University College London, Gower Street, London, WC1E 6BT, UK}

\address[address3]{Complexity Science Hub Vienna, Josefstaedter Strasse 39, A 1080 Vienna}

\begin{abstract}
We describe two different bootstrap methods applied to the detection of a minimum spanning tree obtained from a set of multivariate variables. We show that two different bootstrap procedures provide partly distinct information that can be highly informative about the investigated complex system. Our case study, based on the investigation of daily returns of a portfolio of stocks traded in the US equity markets, shows the degree of robustness and completeness of the information extracted with popular information filtering methods such as the minimum spanning tree and the planar maximally filtered graph. The first method performs a ``row bootstrap" whereas the second method performs a "pair bootstrap". We show that the parallel use of the two methods is suggested especially for complex systems presenting both a nested hierarchical organization together with the presence of global feedback channels.
\end{abstract}


\end{frontmatter}



\section{Introduction}
Multivariate time series or, more generally, multivariate variables are today common in many research areas. Examples are finance, biology, cognitive sciences, medical sciences, economics, physics, social sciences, etc.  In all these research areas is of crucial importance to perform dimensionality reduction, information filtering and visualization of the most relevant information. Many methods have been proposed to perform information filtering from multivariate variables. Examples are dimensionality reduction and information filtering based on principal component analysis, random matrix theory, hierarchical clustering, topological constraints on networks, and statistical validation in network construction.

A simple successful method of information filtering is the extraction of a minimum (or maximum) spanning tree from multivariate variables  \cite{Mantegna1999}. Since 1999, many alternative approaches to the minimum spanning tree have been proposed. Hereafter, we briefly cite only a limited set of alternatives. One of the first alternatives was the one discussed in  \cite{Onnela2002} where a network is built starting from a correlation matrix by inserting links between two nodes when their correlation coefficient is above a given threshold. Other examples consider the clustering procedure achieved by considering Potts super-paramagnetic transitions \cite{Kullmann2000} or maximum likelihood procedure \cite{Giada2001,Giada2002}, where authors define the likelihood by using a one-factor model. An unsupervised and deterministic clustering procedure, labeled as directed bubble hierarchical tree, was proposed in \cite{Song2012}. In spite of the existence of many potential alternatives, the minimum spanning tree (MST) \cite{Mantegna1999}  has been detected and investigated in a large variety of complex systems \cite{Tumminello2010,Marti2017} due to its simplicity and effectiveness.   

Many complex systems present at least part of their structure that is  hierarchically nested \cite{Tumminello2007}. For this reason, different regions of the associated minimum spanning tree are affected to a different degree from the unavoidable uncertainty associated with the empirical estimation of the similarity (or dissimilarity) measure used to perform the information filtering. In ref. \cite{Tumminello2007spanning}, authors investigated the reliability of links of a MST by performing bootstrap replicas of the correlation matrix of a portfolio of stocks when this matrix was used as a similarity matrix. Bootstrap \cite{Efron1994} is a sampling with replacement procedure widely used in non parametric statistical tests.  

In this paper, we investigate two different methods of performing bootstrap replicas of a correlation matrix and we show that the chosen method impacts the information extracted by the MST. The first method is a bootstrap procedure where each time information about the entire system is sampled for all elements (we address this bootstrap sampling procedure as ``row bootstrap"). In the second method the bootstrap estimation of the correlation of a pair of elements is done independently for each pair (we address this bootstrap sampling procedure as ``pair bootstrap").   Our investigation shows that for both cases the filtering of the minimum spanning tree is quite robust with respect to inaccuracies or intrinsic limitations in the estimation of the correlation and that the additional information that can be obtained from the bootstrap analysis is in agreement with the hypothesis that the planar maximally filtered graph \cite{Tumminello2005} includes the most statistically robust information that cannot be present in the minimum spanning tree due to the constraints associated with its tree nature. Our investigation of a portfolio of stock returns also shows that  ``pair bootstrap" and ``row bootstrap" highlight different type of information and that making stocks' partitions by using MST ``pair bootstrap" is systematically performing better than ``row bootstrap" in a classification of stocks done in terms of the economic sectors and economic subsectors of the investigated companies. 

The article is organized as follows. In Sect. \ref{S1} we describe the two distinct bootstrap procedures we use. In Sect. \ref{S2} we present a case study performed by considering a multivariate set of return of 300 stocks traded in a financial market. Sect. \ref{S3} compares the natural partitions of the graph components obtained by using different values of the bootstrap threshold with partitions obtained using experts' classification and in Sect. \ref{S4} we draw our conclusions.

\section{Bootstrap procedures of multivariate variables}
\label{S1}

\begin{figure}
{\includegraphics[width=0.8 \linewidth]{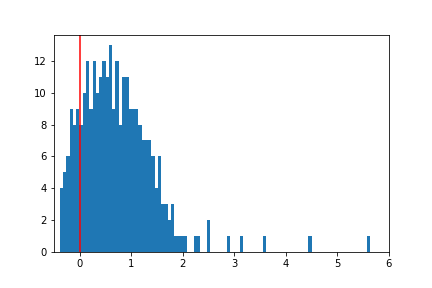}
}
\caption{Eigenvalue spectrum of a correlation matrix obtained with the second method of bootstrap. The largest eigenvalue is out of the $x$ scale. A certain number of negative eigenvalues are observed in the interval $-0.3 < \lambda < 0$ of are present.
}
\label{FigB}
\end{figure}

Multivariate variables monitor a complex system by  describing the $n$ elements of the systems with $k$ attributes for each of them. In the present work we will apply our method to a system of $n$ stocks each one characterized by $T$ daily records of the stock returns. Specifically, as illustrative example we investigate the 300 most capitalized stocks traded at US equity markets during the calendar years from 2001 to 2003. In our analyses, we use daily return time series computed as the logarithm of the ratio of closure price to open price. The number of daily records is $T = 748$. Stocks are classified according to their economic sector and sub-sector by using the classification scheme of Yahoo finance.
\begin{figure}
{\includegraphics[width=0.8 \linewidth]{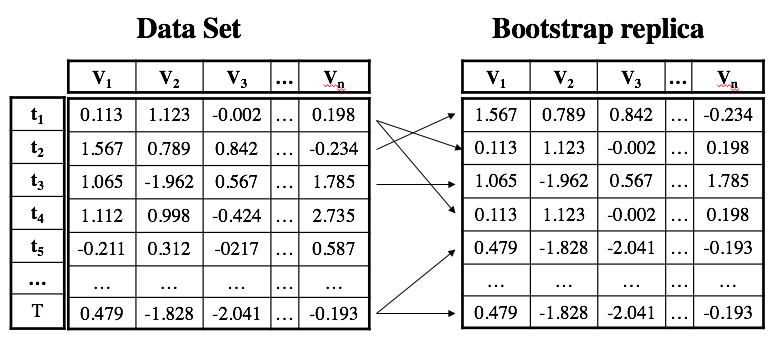}
\includegraphics[width=0.5 \linewidth]{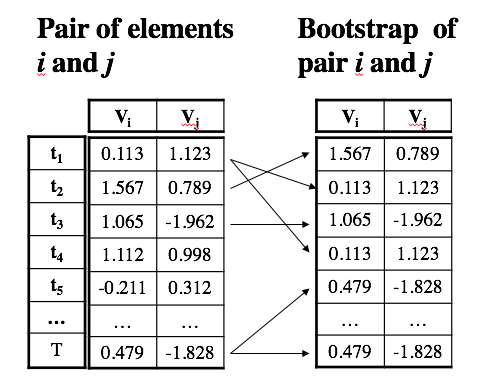}
}
\caption{\scriptsize{Top: Schematic description of the``row bootstrap". Rows of different time records are sampled with replacement and the bootstrap replica is obtained. The correlation matrix is then computed from the bootstrap replica of vector of data. Bottom:  Schematic description of the``pair bootstrap". For each pair of elements $i$ and $j$ synchronous records are sampled with replacement and the correlation coefficient $\rho(i,j)$ is computed. The procedure is repeated for all pairs and the correlation matrix is obtained.
}}
\label{FigA}
\end{figure}
In Fig. \ref{FigA} we illustrate the basic aspects of the two bootstrapping methods used in this paper to obtain bootstrap replicas of the empirical correlation matrix. In the top panel of Fig. \ref{FigA} we present a schematic description of the``row bootstrap". Fig. \ref{FigA} shows that with this method rows of different time records are sampled with replacement and a bootstrap replica of the $n \times T$ matrix of data is obtained. Starting from the bootstrap replica of the data the correlation matrix is then computed. In the bottom panel we illustrate the``pair bootstrap". In this case the estimation of the correlation coefficient is done for each pair independently. In fact, for each pair of elements $i$ and $j$ synchronous records are sampled with replacement and the correlation coefficient $\rho(i,j)$ is computed. The procedure is repeated for all pairs and the correlation matrix is obtained. It is worth noting that method one provides a well defined correlation matrix, i.e. a positive definite correlation matrix, by construction whereas the second method does not guarantee that the replica correlation matrix is positive definite due to statistical fluctuations in the estimation of the correlation coefficients. We have verified that the correlation matrix obtained by ``pair bootstrap" has associated an eigenvalue spectrum presenting some negative eigenvalues (see Fig \ref{FigB} for an example of the eigenvalue spectrum of a correlation matrix obtained with the second method of bootstrap). In case the positive definite property of the correlation matrix is strictly required, this limitation can be overcomed by performing a shrinkage to some model correlation matrix.    

\section{Case study}
\label{S2}
\begin{figure}
{\includegraphics[width=0.8 \linewidth]{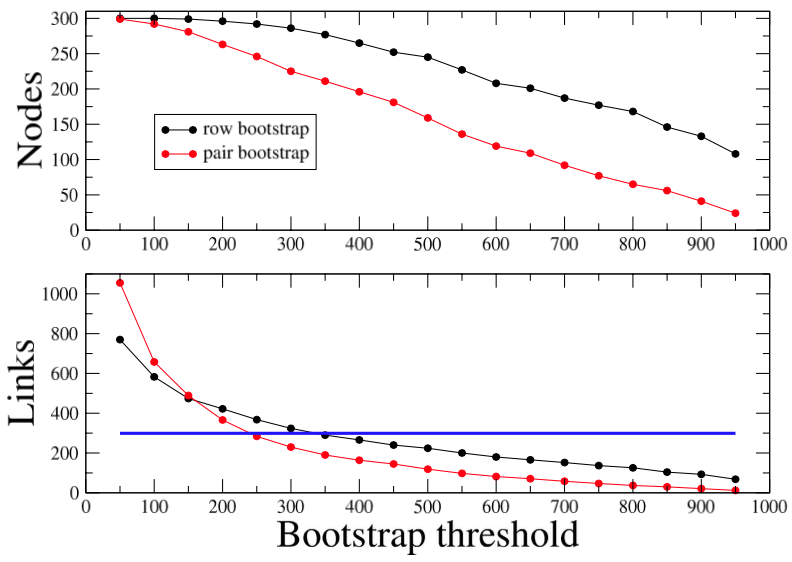}
}
\caption{Number of nodes (top panel) and number of links (bottom panel) of the networks obtained by selecting links that are characterized by a bootstrap value higher than the bootstrap threshold. For each method, ``row bootstrap" method (black symbols) and ``pair bootstrap" method (red symbols) we perform 1,000 bootstrap replicas. The blue line indicates the number of links present in the MST.}
\label{FigD}
\end{figure}

In the present section we present the results of our investigation of the MSTs obtained by performing 1,000 bootstrap replicas of the data described in Sec. \ref{S1} with both the ``row bootstrap" and the ``pair bootstrap" method. For both methods, we compute 1,000 bootstrap replicas of the correlation matrix of stock returns and for each replica we estimate the corresponding MST. For each MST we consider the links that are present in the different bootstrap replicas. The ``row bootstrap" method detects 3357 links (out of 44850 possible distinct links) that are observed in the bootstrap replicas at least once. We address as bootstrap value the number of time that a specific link is present in the 1,000 bootstrap replicas. For the ``row bootstrap" method we detect a histogram of bootstrap value of links that is ranging from 1 to 1,000.  The same interval of bootstrap value is also observed for the  ``pair bootstrap" method that detects 11138 links (out of 44850 possible distinct links).  
\begin{figure}
{\includegraphics[width=0.8 \linewidth]{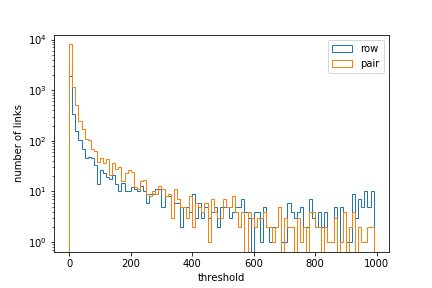}
}
\caption{Histogram of the number of links with given bootstrap value for the links of MSTs of 1,000 bootstrap replicas obtained with    the ``row bootstrap" method (blue line) and the ``pair bootstrap" method (orange line).
}
\label{FigC}
\end{figure}
In Fig. \ref{FigC} we show the histogram of the number of links with a given bootstrap value both for the ``row bootstrap" method and for the ``pair bootstrap" method. The figure shows that in both cases links are characterized by bootstrap values ranging from 1 to 1,000. As expected, for lower values of the bootstrap value the number of links is increasing especially for a bootstrap value of the order of 200 or less. By selecting a specific value of the bootstrap threshold we can therefore filter those links that are observed in the MSTs of bootstrap replicas that are characterized by a statistical robustness of the order of higher than the bootstrap threshold. When we use a high value of the bootstrap threshold we select an information which is highly stable with respect to sampling the real data with replacement but this information can cover only part of the investigated system. Fig. \ref{FigD} summarizes the information concerning the number of nodes and links observed in the networks obtained by selecting links that are characterized by a bootstrap value higher than the bootstrap threshold.   
The figure shows that ``row bootstrap" method (black symbols) and ``pair bootstrap" method (red symbols) are selecting networks with a different number of nodes and a different number of links for each bootstrap threshold. Specifically, the networks obtained with the ``row bootstrap" method always have a larger number of nodes than the networks obtained with the ``pair bootstrap" method at a given bootstrap threshold. The behavior of the number of links as a function of the bootstrap threshold is more variated. Specifically, networks selected with the ``row bootstrap" methods have more links than the other ones for values of the bootstrap threshold higher than 150 whereas the reverse is true for lower values. Fig. \ref{FigD} also shows that the number of links of the selected networks is less than the number of links expected for the MST (which is $n-1=299$) for a bootstrap value higher than $\approx 350$ for  ``row bootstrap" and  higher than $\approx 250$ for  ``pair bootstrap" indicating that the MST filtering is quite robust in selecting links which are highly stable with respect to bootstrap validation. In fact, only when the bootstrap threshold assumes relatively low values a number of links significantly higher than $n-1$ is observed and indeed this number never exceeds values higher than $4n$.  

\begin{figure}
{\includegraphics[width=0.8 \linewidth]{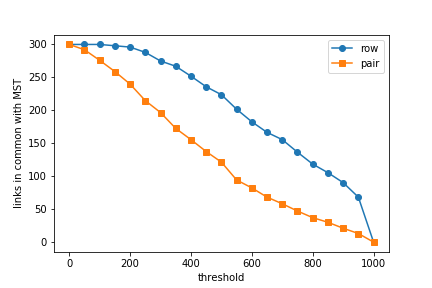}
}
\caption{Number of links in common between the networks obtained by setting a given bootstrap threshold for the two bootstrap methods of ``row bootstrap" (blue symbols) and ``pair bootstrap" (orange symbols). The original MST has $n-1=299$ links.}
\label{FigE}
\end{figure}
The links selected by the bootstrap procedure overlap with the MST obtained from original data. In Fig. \ref{FigE} we show the number of links in common between the networks obtained by setting a given bootstrap threshold with the two different methods and the original MST. The figure shows that the inclusion of the original MST is almost complete when the ``row bootstrap" method is used already for bootstrap threshold of the order of 200 or less. The inclusion is less pronounced for the ``pair bootstrap" method. In fact for this method bootstrap threshold as low as 50 need to be used to achieve an almost total inclusion.

For bootstrap threshold as low as approximately $150$,  the two bootstrap methods are confirming the informational role of the majority of links of the original MST and are also highlighting other links that are not selected in the original MST due to the severe topological constraint of this network that it is requiring that the selected links are forming a tree. This intrinsic limitation of the MST is known and attempts have been done to propose extensions of it that are including additional information present in the correlation matrix. One direct extension of the MST is the planar maximally filtered graph (PMFG) obtained starting from a proximity (similarity or dissimilarity) measure by imposing a topological constraint on the surface embedding the network \cite{Tumminello2005}. The PMFG always includes the MST but has more links and also includes cliques of 3 and 4 nodes. 

\begin{table}
\caption{Number of 3-cliques observed in the networks obtained with the two bootstrap methods 
of ``row bootstrap" (RB) and ``pair bootstrap" (PB). Columns 4 and 5 show the number of these cliques that are present also in the original PMFG. Columns 6 and 7 report the percent of inclusion of the 3-cliques in the original PMFG. No 3-clique is observed for bootstrap threshold higher or equal to 550.}
\label{TabA}
 \begin{tabular}{lrrrrrr}
\hline
\hline
Bootstrap & RB	& PB & RB & PB & RB \% & PB \% \\
threshold       &  number of & number of & 3-cliques & 3-cliques & 3-cliques & 3-cliques \\
~~       &  3-cliques 	& 3-cliques & in PMFG & in PMFG & in PMFG & in PMFG \\
\hline
550	& 0 & 0 & 0 & 0 & -- & -- \\
500  & 1 & 0 & 1 & 0 & 100 & -- \\
450  & 4 & 2 & 4 & 2 & 100 & 100 \\
400  & 12 & 4 & 12 & 4 & 100 & 100 \\
350  & 20 & 12 & 20 & 12 & 100 & 100 \\
300  & 32 & 26 & 32 & 26 & 100 & 100 \\
250  & 57 & 57 & 55 & 57 & 96.5 & 100 \\
200  & 91 & 122 & 84 & 111 & 92.3 & 91.0 \\
150  &128 & 253 & 116 & 209 & 90.6 & 82.6 \\
100 	& 245 & 512 & 183 & 330 & 74.7 & 64.4 \\
50    & 470 & 1319 & 282 & 489 & 60.0 & 37.1 \\
\hline
\end{tabular}
\end{table}

In Table \ref{TabA}  we report the number of 3-cliques observed in the networks obtained with the two bootstrap methods 
of ``row bootstrap" (RB) and ``pair bootstrap" (PB). The networks are always trees for bootstrap values higher or equal to 550. The first 3-clique is observed for a bootstrap threshold equal to 500. Starting from this value, the number of 3-cliques increases by decreasing the bootstrap threshold. It is worth noting that for values of the bootstrap threshold higher or equal to 150 more than 80\% of these 3-cliques are included in the PMFG of original data. This observation support the conclusion that PMFG includes links that are highly informative but that cannot be included in the MST due to the severe topological constraint requiring this network to be a tree. 

We have no theoretical or numerical arguments to set a specific value of the bootstrap threshold. However, the analysis of the results presented so far suggests that a bootstrap threshold higher or equal to 150 is highly reliable and informative for this system.

\section{Comparison of obtained partitions with an experts' classification}
\label{S3}
The networks obtained with the two bootstrap methods at different values of the bootstrap threshold are typically presenting several disconnected components. This characteristic is very pronounced for high values of the bootstrap threshold and it progressively disappears by decreasing the bootstrap threshold. To assess the relative degree of informativeness of the two bootstrap methods when describing a subset of the investigated system we have compared the partitioning observed in the obtained networks with a classification of the stocks done by experts. Specifically, we measure the adjusted Rand index (ARI) \cite{Hubert1985} between the partition observed in a network characterized by a given bootstrap threshold and involving a given number of (non isolated) nodes with the partition produced by financial experts classifying each stock in terms of an economic sector and of an economic subsector. 
\begin{figure}
{\includegraphics[width=1.0 \linewidth]{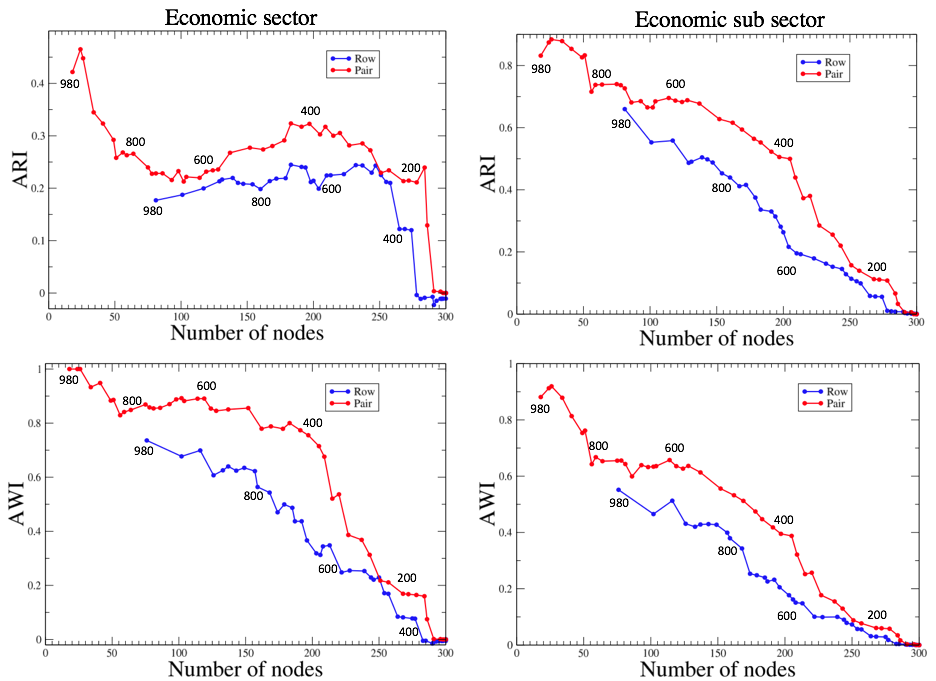}
}
\caption{ARI (top panels) and AWI (bottom panels) of the networks obtained with the two bootstrap methods of ``row bootstrap" (blue symbols) and ``pair bootstrap" (red symbols) with the partition obtained using information about economic sectors (left panels) of stocks or economic sub-sectors of stocks (right panels). Each point represent a value of the bootstrap threshold. Bootstrap threshold is ranging from 980 (the starting point on the left) to 20 (the final point on the right) with step of 20. Some values of the Bootstrap threshold are written close to the corresponding point. }
\label{FigF}
\end{figure}

The results of this comparison are summarized in Fig. \ref{FigF} where we show the ARI and the adjusted Wallace index (AWI) between the partitions observed in the networks obtained with bootstrap methods at different bootstrap thresholds (indicated in the figure as numbers near a specific point)  and the partitions obtained from experts' classification at the level of economic sector (left panels) and economic sub sectors (right panels). When two memberships are compared pairwise the precision in observing a pair in the same partition is usually addressed as one of the Wallace indices \cite{Wallace1983}. Ref. \cite{Bongiorno2017} has generalized the Wallace index to obtain an adjusted index varying  between $-\infty$ and one. It assumes a zero value when the test partition is included into the reference partition not better than in a random case.  A high value of the AWI indicates a high precision in classifying pairs of nodes that are belonging to the same community as defined in the reference partition.  

The panels of Fig. \ref{FigF} show that the number of stocks included into the  network is increasing  by decreasing the bootstrap threshold. The ARI observed with the classification based on the economic sectors (right top panel) has a non monotonic profile showing a local increase for number of nodes of the order of 200 for "pair bootstrap" and 230 for "row bootstrap" while in the case of economic sub-sectors (left top panel) a global monotonic decrease is observed. The observed decrease is associated with the progressive interconnection of the network. In fact nodes are progressively connected among them when the bootstrap threshold decreases and the presence of a single connected component has associated a zero value for the ARI with the partitioning done in terms of experts' classification. In other words the decrease of the ARI is expected as function of the number of not isolated nodes present in the network. What is not expected is the observation that the ARI of the "pair bootstrap" method always exceeds the ARI of the "row bootstrap" methods when we compare two networks characterized by the same number of non isolated nodes. It should be noted that the two networks are obtained by using two different bootstrap thresholds and the bootstrap threshold used by the "pair bootstrap" is lower, i.e. might in principle select links with less precision, that the one used by the "row bootstrap".
The same conclusion is reached when we analyze the behavior of the AWI as a function of the number of not isolated nodes of the networks obtained with the bootstrap methods. When the number of nodes is fixed, the "pair bootstrap" method has higher value of the AWI than the "row bootstrap method".
\begin{figure}
{\includegraphics[width=1.0 \linewidth]{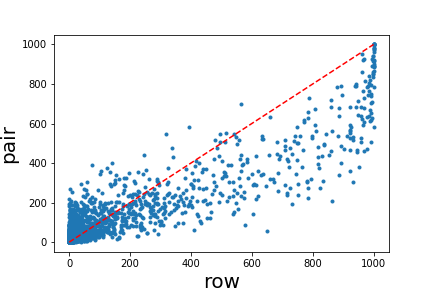}
}
\caption{Scatter plot of the bootstrap value of each link detected at least once with at least one of the two bootstrap methods. The red dashed line is the $y=x$ line. }
\label{FigG}
\end{figure}

The results we have seen so far show that the two methods work differently the one from the other. It is therefore of interest to assess what are the main differences between the two methods. In Fig. \ref{FigG} we show the scatter plot of the bootstrap value of each link detected at least once with at least one of the two bootstrap methods. The links with bootstrap value greater than zero in both methods are 3280. Both links with the "row" bootstrap value higher than "pair" bootstrap value and vice versa are observed. The two methods are highlighting different type of information. Specifically, the "row bootstrap" takes into account the global market movement and usually presents higher values of the bootstrap values  for links of higher correlation coefficient. The ``pair bootstrap" method focuses on each specific pair without putting a special emphasis on the most pronounced global market movements. Due to this it estimates the correlation of the pair without relying too much on events that have affected the entire market.

By performing a statistical test assuming as a null hypothesis that links presenting "row" bootstrap value $r_{bv}$ larger than the  "pair" bootstrap value $p_{bv}$ are randomly distributed with respect to the two binary states characterized by (0) same economic sector of the two nodes linked, and (1) different economic sector of the two nodes linked. Our statistical test shows that the null hypothesis is rejected and that links characterized by the condition $r_{bv} > p_{bv}$ are preferentially having nodes belonging to different economic sector while the opposite is true when the condition $p_{bv} > r_{bv}$ holds. In other words, the "row bootstrap" put emphasis on those time records with market events affecting a large set of stocks whereas the "pair bootstrap" is more sensitive to the comovements of the pair of stocks that are distributed in many records and not necessarily affecting synchronously a large number of stocks.     

\section{Conclusions}
\label{S4}
In this paper we show that two different bootstrap procedures provide information about a set of multivariate variables that can be highly informative with respect to the information present in the proximity (in our case a distance obtained starting from a correlation measure) matrix and to the robustness and completeness of the information extracted with popular information filtering methods such as the MST and the PMFG. The first method performs a ``row bootstrap" whereas the second method performs a "pair bootstrap". The first method provides bootstrap replicas of the data associated with a positive definite correlation matrix whereas the second one cannot guarantee this property but provides networks where information about the pair interaction is obtained without relying implicitly on the impact of records that are affecting globally the system. The "row bootstrap" method is faster than the "pair bootstrap" method but the two methods are not carrying the same information. For this reason, especially for complex systems presenting both a nested hierarchical organization together with the presence of global feedback channels  it might be informative to perform both methods and evaluate the different type of information that can be extracted by each of them.      

{\bf{Acknowledgments}}
Authors wish to thank participants of the Econophysics Colloquium 2017 and especially Tomasz Gubiec for comments and  feedback provided at the Colloquium after the presentation of this work.

\bibliography{MMMM2}

\end{document}